\documentclass[12pt,number,sort&compress]{elsarticle}

\usepackage[margin=1in]{geometry}
\usepackage{caption}
\usepackage{subcaption}
\usepackage{amsmath}
\usepackage{amssymb}
\usepackage[numbers]{natbib}
\usepackage[version=3]{mhchem}
\usepackage{multirow}
\usepackage{xr}
\usepackage{xcolor}
\usepackage[normalem]{ulem}
\usepackage{mciteplus}
\usepackage[colorlinks=true,linkcolor=black, citecolor=blue, urlcolor=blue]{hyperref}
\definecolor{darkgreen}{rgb}{0.0,0.5,0.0}
\mciteErrorOnUnknownfalse
\usepackage{float}
\usepackage{bm}
\usepackage{longtable}
\usepackage{multirow}
\usepackage{array}

\newcolumntype{P}[1]{>{\centering\arraybackslash}p{#1}}
\newcolumntype{M}[1]{>{\centering\arraybackslash}m{#1}}

\makeatletter
\newcommand*{\addFileDependency}[1]{
  \typeout{(#1)}
  \@addtofilelist{#1}
  \IfFileExists{#1}{}{\typeout{No file #1.}}
}
\makeatother

\begin{document}
\begin{frontmatter}
\title{Complex strengthening mechanisms in nanocrystalline Ni-Mo alloys revealed by a machine-learning interatomic potential}
\author[SYSU]{Xiang-Guo Li \corref{cor1}}
\ead{lixguo@mai.sysu.edu.cn}

\author[OU]{Shuozhi Xu}
\author[SYSU]{Qian Zhang}
\author[SYSU]{Shenghua Liu}
\author[SYSU]{Jing Shuai\corref{cor1}}
\ead{shuaij3@mail.sysu.edu.cn}

\cortext[cor1]{Corresponding author}

\address[SYSU]{School of Materials, Shenzhen Campus of Sun Yat-sen University, No.\ 66, Gongchang Road, Guangming District, Shenzhen, Guangdong 518107, PR China}
\address[OU]{School of Aerospace and Mechanical Engineering, University of Oklahoma, Norman, OK 73019-1052, USA}

\begin{abstract}
A nanocrystalline metal's strength increases significantly as its grain size decreases, a phenomenon known as the Hall-Petch relation. Such relation, however, breaks down when the grains become too small. Experimental studies have circumvented this problem in a set of Ni-Mo alloys by stabilizing the grain boundaries (GB). Here, using atomistic simulations with a machine learning-based interatomic potential (ML-IAP), we demonstrate that the inverse Hall-Petch relation can be correctly reproduced due to a change in the dominant deformation mechanism as the grain becomes small in the Ni-Mo polycrystals. It is found that the atomic von Mises strain can be significantly reduced by either solute doping and/or annealing for small-grain-size polycrystals, leading to the increased strength of the polycrystals. On the other hand, for large-grain-size polycrystals, annealing weakens the material due to the large atomic movements in GB regions. Over a broad range of grain size, the superposition of the solute and annealing effects on polycrystals enhances the strength of those with small grain size more than those with large ones, giving rise to the continuous strengthening at extremely small grain sizes. Overall, this study not only demonstrates the reliability of the ML-IAP, but also provides a comprehensive atomistic view of complex strengthening mechanisms in nanocrystals, opening up a new avenue to tailor their mechanical properties.
\end{abstract}

\end{frontmatter}

\section{Introduction}

The strength of polycrystalline materials is known to increase with the reduction of the grain size \cite{cheng2005tensile,yin2018ultrastrong}, called the Hall-Petch relation. One possible strengthening mechanism is based on dislocation pileups at grain boundaries (GBs), which hinds dislocation motion \cite{xu_comparing_2017}. However, the dislocation-based deformation mechanism transits to GB-dominated plasticity at extremely fine grain sizes \cite{schiotz1998softening,schiotz2003maximum,shen2007effect}, leading to an inverse Hall-Petch relation. Solute segregation is another strategy to increase the material strength \cite{pan2020heterogeneous,li2021role,hu2017grain,leyson2010quantitative}. Solute-dislocation interaction is believed to be a key factor for the solute strengthening effect \cite{li2021role,varvenne2017solute}. Pan and Sansoz \cite{pan2020heterogeneous} recently found that the solute clusters from heterogeneous solute segregation can suppress strain localization, and are responsible for solute strengthening. Moreover, Hu {\textit{et al}} \cite{hu2017grain} have demonstrated that Mo segregation can stabilize GBs in nanocrystalline Ni-Mo alloys, leading to an increased strength and a resurgence of Hall-Petch strengthening with grain size down to a few nanometers. Experimental results \cite{hu2017grain} also show that annealing can enhance the Hall-Petch strengthening. However, the atomistic strengthening mechanisms, particularly regarding the coupling effects of solute, annealing, and grain sizes, still need clarification because they are difficult to uncover in experiments \cite{hu2017grain}.

Due to the high computational cost, the density functional theory (DFT) calculations have been limited to several hundreds of atoms. Therefore, higher-scale computational tools such as molecular dynamics (MD) have become an important approach to study mechanical properties and the associated underlying mechanisms \cite{li2010dislocation,yamakov2002dislocation,zhang2017formation,antillon2019molecular} in metals and alloys. Atomistic simulations using linear-scaling interatomic potentials can potentially access large systems and long timescales. The accuracy of the atomistic simulations largely depends on the interatomic potentials used \cite{chavoshi_addressing_2017}. Li {\textit{et al}} \cite{li2019regulating} reported a mechanism transition from GB-accommodated plastic deformation to dislocation-based plasticity below the optimal size for the maximum strength after segregating Mo solute atoms in Ni-Mo alloys using MD simulations with an embedded atom method (EAM) potential. Sansoz {\textit{et al}} \cite{sansoz2022hall} recently used an EAM potential to systematically study the effects of solute concentration on the Hall-Petch strengthening limits with hybrid Monte-Carlo/MD (MC/MD) simulations in Ag-Cu alloys. The EAM potential, and broadly most classical interatomic potentials, are fitted mainly to elemental properties and thus generally perform poorly for alloys. Recently, the development of machine learning-based interatomic potential (ML-IAP) \cite{behler2007generalized,dragoni2018achieving,thompson2015spectral,shapeev2016moment,zuo2020performance} provides another possibility, which can reach near-DFT accuracy at several of orders magnitude lower cost than DFT. In the last few years, ML-IAP has been extensively applied to revealing the contributing factors of alloys' mechanical properties, including the lattice distortion \cite{kostiuchenko2019impact,jafary2019applying}, short-range ordering \cite{yin2021atomistic,li2020complex}, defect and dislocation properties \cite{maresca2018screw,goryaeva2021efficient}, etc. To the best of our knowledge, prediction of the inverse Hall-Petch relation by ML-IAP has not been realized yet. In addition, although plenty of theoretical studies have been performed to study different strengthening mechanisms, the coupling between these mechanisms, e.g. coupling between grain size strengthening, solute and annealing effects, remains elusive. 

In this work, we utilize our previously developed spectral neighbor analysis potential (SNAP) \cite{li2018quantum} to investigate the grain size, solute, and annealing effects in Ni-Mo polycrystalline systems. We demonstrate that the ML-IAP can accurately predict the inverse Hall-Petch relation and reveal the different plasticity mechanisms with the dominant role involving GB or dislocation at different grain sizes. Our results indicate that both solute doping and annealing can reduce the atomic von Mises strain of the polycrystals at yielding under uniaxial tensile strain, stabilize the GB, and thus increase the strength of the polycrystals at small grain sizes, leading to the resurgence of the Hall-Petch strengthening at grain sizes within 10 nanometers. For large grain-size polycrystals, solute doping can increase the dislocation density giving rise to the enhancement of the polycrstal strength, while annealing, on the other hand, would induce sizable atomic strain at the GB during plastic flow deformation, leading to a decrease in the strength of the polycrystals.

\section{Method}
\subsection{Polycrystal model setup}

\begin{figure}[h]
\includegraphics[width=1.0\textwidth]{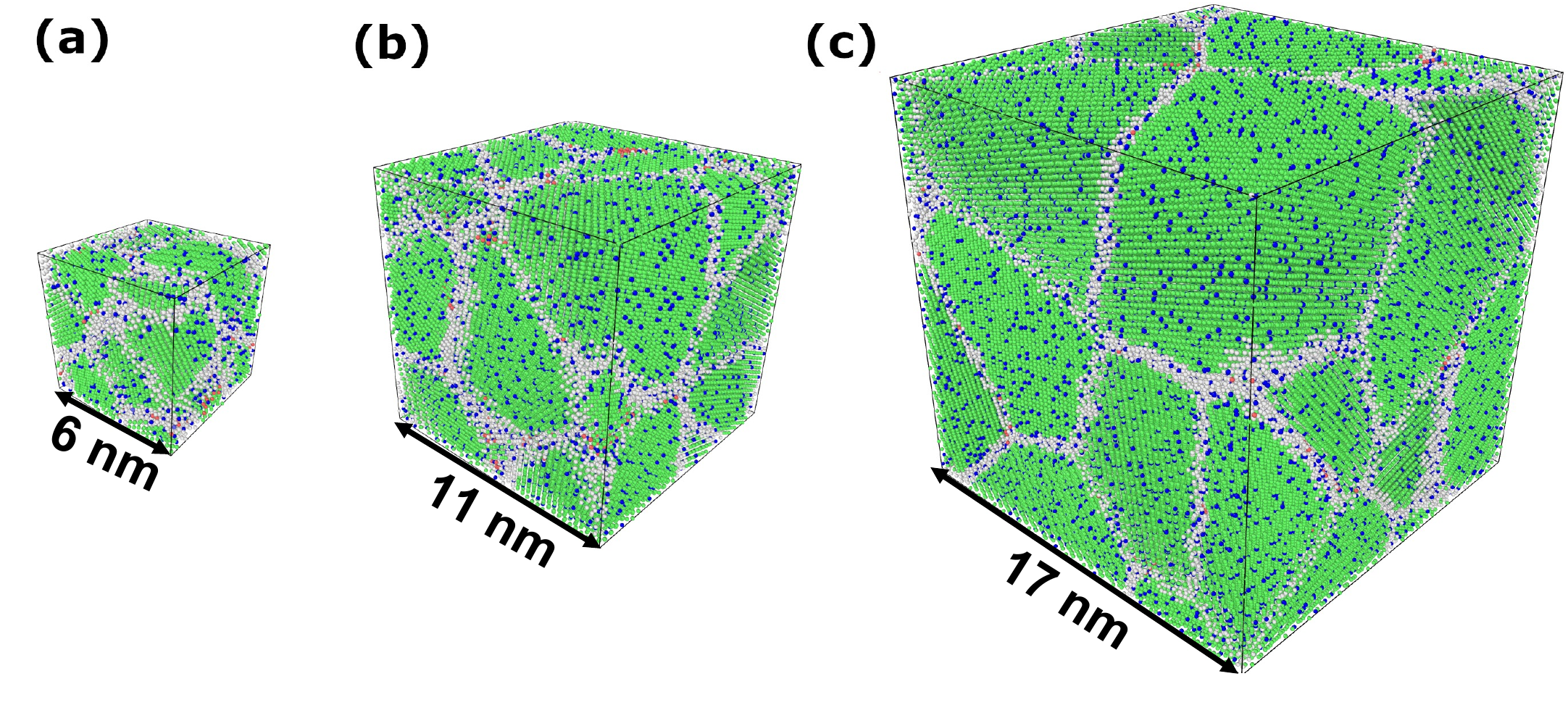}
\caption{\label{fig:str}The polycrystalline models with different grain sizes. Configurations of nanostructured alloys with edge lengths of (a) 6 nm, (b) 11 nm, and (c) 17 nm and average grain size of (a) 4.1 nm, (b) 7.5 nm, and (c) 11.6 nm. Green spheres are atoms in local FCC structures, while white spheres are atoms in disordered structures, i.e., atoms within the GB regions. All atoms are Ni, except those blue ones which are Mo.}
\end{figure}

We generated the initial Ni polycrystal models using the Voronoi tessellation method \cite{brostowConstructionVoronoiPolyhedra1978} implemented in the Atomsk \cite{hirelAtomskToolManipulating2015} code. A number of cubic supercells were constructed with different side lengths, six grains were then randomly inserted giving rise to a series of polycrystals with different average grain diameters. We present six polycrystals with average grain diameters of 4.1, 6.1, 7.5, 8.8, 10.2 and 11.6 nm and corresponding edge lengths of 6, 9, 11, 13, 15, 17 nm, respectively. Periodic boundary conditions are imposed on all three dimensions. Neighboring atoms with a distance $<1.5$ \AA ~were removed at the GBs. The number of atoms in the polycrystals ranges from $\sim 2,000$ to $\sim 454,600$. The Ni-Mo polycrystalline models were constructed by randomly replacing 10$\%$ of Ni atoms with Mo. This percentage is lower than the limit of solubility of Mo in Ni \cite{hu2017grain}. Three atomistic models with different average grain diameters are shown in Fig.~\ref{fig:str}.

\subsection{Interatomic potential}

We performed the hybrid MD and Monte Carlo (MC) calculations using the large-scale atomic/molecular massively parallel simulator (LAMMPS) package \cite{plimptonFastParallelAlgorithms1995} with the Ni-Mo SNAP model \cite{li2018quantum}, previously developed by the lead author. The training structures of this ML-IAP include 1) Ni, Mo, Ni$_3$Mo compound, Ni$_4$Mo compound, and their distorted structures, 2) surface structures of Ni and Mo, 3) snapshots of {\textit{ab initio}} molecular dynamics (AIMD) simulations at different temperatures and distortions, 4) alloy structures constructed from partial substitution. The reliability of this potential has been checked and validated with the following properties: 1) lattice constants, 2) surface energies, 3) elastic constants, 4) defect properties, 5) melting points, 6) phase diagram. A good agreement within these properties between the results from the SNAP Ni-Mo model and that from DFT/experiment has been achieved \cite{li2018quantum}.

\subsection{Annealing and tensile deformation}
The polycrystals were thermally equilibrated at 300 K for 0.1 ns for pure Ni polycrystal and 0.4 ns for Ni-Mo polycrystal, via MD (for pure Ni polycrystal) and MC/MD (for Ni-Mo polycrystal with Mo solutes) simulations, respectively, in an isothermal-isobaric NPT ensemble. For annealing, the polycrystal will be further annealed at annealing temperature (600 K) for 0.75 ns, and then quenched from the annealing temperature to room temperature in 0.15 ns, followed by another equilibrium at room temperature for 50 ps. Uniaxial tensile deformation was then applied in the $z$-direction at a strain rate of 5$\times 10^8$ s$^{-1}$ for 0.2 ns at 300 K. To maintain zero lateral pressure (constant uniaxial strain rate), we use NPT ensemble in the $x$- and $y$-directions during the deformation. The time step was set to 1 fs. We use OVITO \cite{stukowskiVisualizationAnalysisAtomistic2009} to visualize the atomic configurations and analyze
simulation results by identifying phase structures (common neighbor analysis \cite{honeycuttMolecularDynamicsStudy1987}) and calculating the atomic strains \cite{falk1998dynamics} from two atomic configurations, a current configuration (deformed one) and a reference configuration (initial one).  To capture the randomness in the distribution of GBs and Mo atoms, we performed three simulations with different initial polycrystalline structures for each grain size.

\section{Results}

\subsection{Hall-Petch and inverse Hall-Petch relation}
\begin{figure}[t]
\includegraphics[width=1.0 \textwidth]{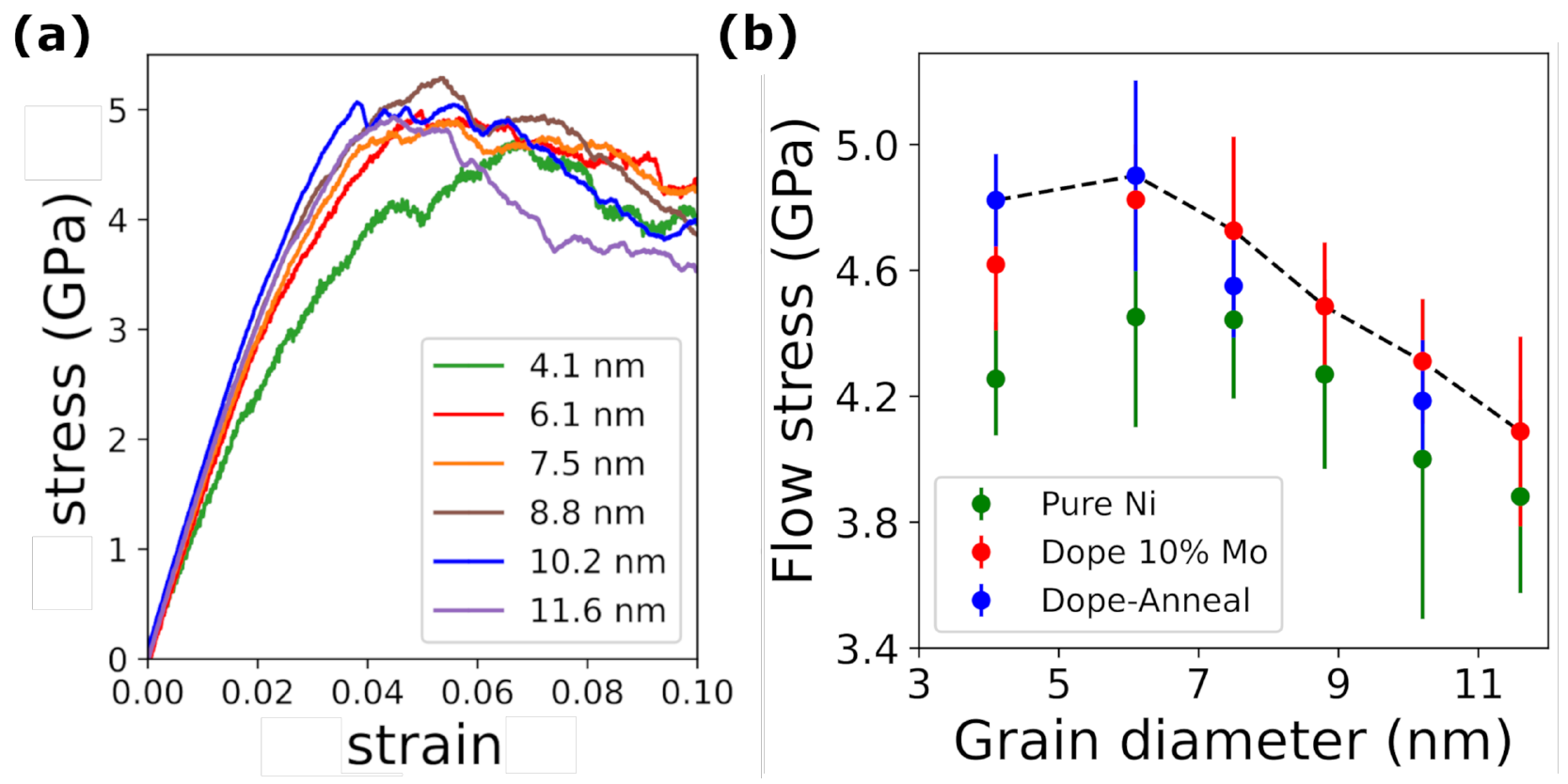}
\caption{\label{fig:hallpetch}The grain-size dependence of the flow stress. (a) Stress-strain curves for three simulations with average grain sizes ranging from 4.1 nm to 11.6 nm for Ni polycrystals. (b) The flow stress of pure Ni polycrystal (green) and Ni-Mo polycrystal with 10$\%$ Mo doping (red), defined as the average stress in the strain interval from 8 to 10$\%$ deformation. The error bars indicate the standard deviation of the three calculations with different initial polycrystalline structures. A maximum in the flow stress is seen for the grain size of around 7 nm for clean Ni and slightly left shift (smaller grain size) for Ni-Mo. The blue point is the flow stress for Ni-Mo polycrystalafter annealing at 600 K. The dark dashed line is guided for eyes for continuous Hall-Petch strengthening at even smaller grains after stabilizing the nano-polycrystals.}
\end{figure}

To test the Hall-Petch and inverse Hall-Petch relations, we plotted the stress-strain curves with varying grain sizes of Ni polycrystal, as shown in Fig.~\ref{fig:hallpetch}a. Since we start from perfect dislocation-free structures and the dislocations need to be nucleated at a large stress from GBs, no dislocations are observed at small strains. As the strain increases to around 4$\%$, the dislocations appear and massive plastic deformation occurs, corresponding to a stress dropping. When the strain surpasses about 8$\%$, the stress becomes more steady. Hence, we calculate the average stress in the strain interval from 8 to 10\%, and take it as the flow stress. Fig.~\ref{fig:hallpetch}b shows that the flow stress depends strongly on the grain size. Specifically, as the grain size decreases from 11.6 nm, the flow stress first increases, and after reaching the maximum (at the grain size of around 7~nm), it decreases. This is the well-known Hall-Petch and inverse Hall-Petch relations \cite{schiotz1998softening,schiotz2003maximum,shen2007effect}. 

\begin{figure}[t]
\includegraphics[width=1.0 \textwidth]{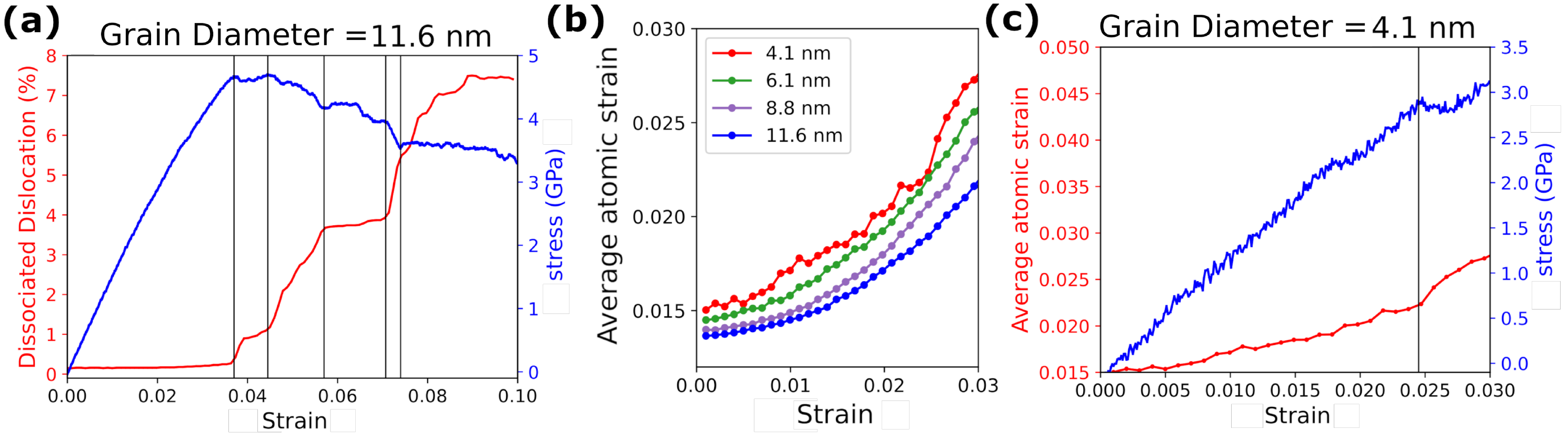}
\caption{\label{fig:analysis}The evolution of dissociated dislocation, stress, and atomic strain as a function of the tensile strain in the Ni polycrystals. (a) Stress-strain and dissociated dislocation-strain curves for polycrystal with large grain size (grain diameter $= 11.6$ nm). (b) The average atomic strain as a function of tensile strain for different grain sizes. (c) Stress-strain and average atomic strain-strain curves for Ni polycrystal with small grain size (grain diameter$=$4.1 nm). The black vertical lines indicate the locations with abrupt changes and guide for eyes.}
\end{figure}

When the grain size is large (e.g. diameter $\geq 8$ nm in our simulations), the flow stress increases with a decreasing grain size, which is consistent with the Hall-Petch relation. This is because, at large grain sizes, dislocation-based deformation dominates, which is indicated by our stress-strain and dissociated dislocation-strain curves in Fig.~\ref{fig:analysis}a. The evolution of the strength of the stress is strongly correlated with the dissociated dislocation. As the grain size further decreases, the GB-mediated plasticity plays a more important role. We plotted the average atomic strain at yielding (from 0 to 3$\%$ applied strain) as a function of the strain for different grain sizes, as shown in Fig.\ref{fig:analysis}b. Noting that at small applied strain, the induced atomic strain is mainly contributed from the GB region. We can see that smaller grain size will induce larger atomic strain under small tensile strain, leading to the decrease of the flow stress. In particular, when the grain size is small (e.g. the grain diameter$=$4.1 nm), we observe an abrupt increase in the atomic strain around 2.5$\%$ strain in the average atomic strain-tensile strain curve. This abrupt change can exactly be identified in the stress-strain curve with obvious plasticity, as shown in Fig.~\ref{fig:analysis}c, which indicates that the GB-mediated plasticity dominates for polycrystals at small grain sizes and further demonstrates the validity of the SNAP model.

\subsection{Solute strengthening}

\begin{figure}[t]
\includegraphics[width=1.0 \textwidth]{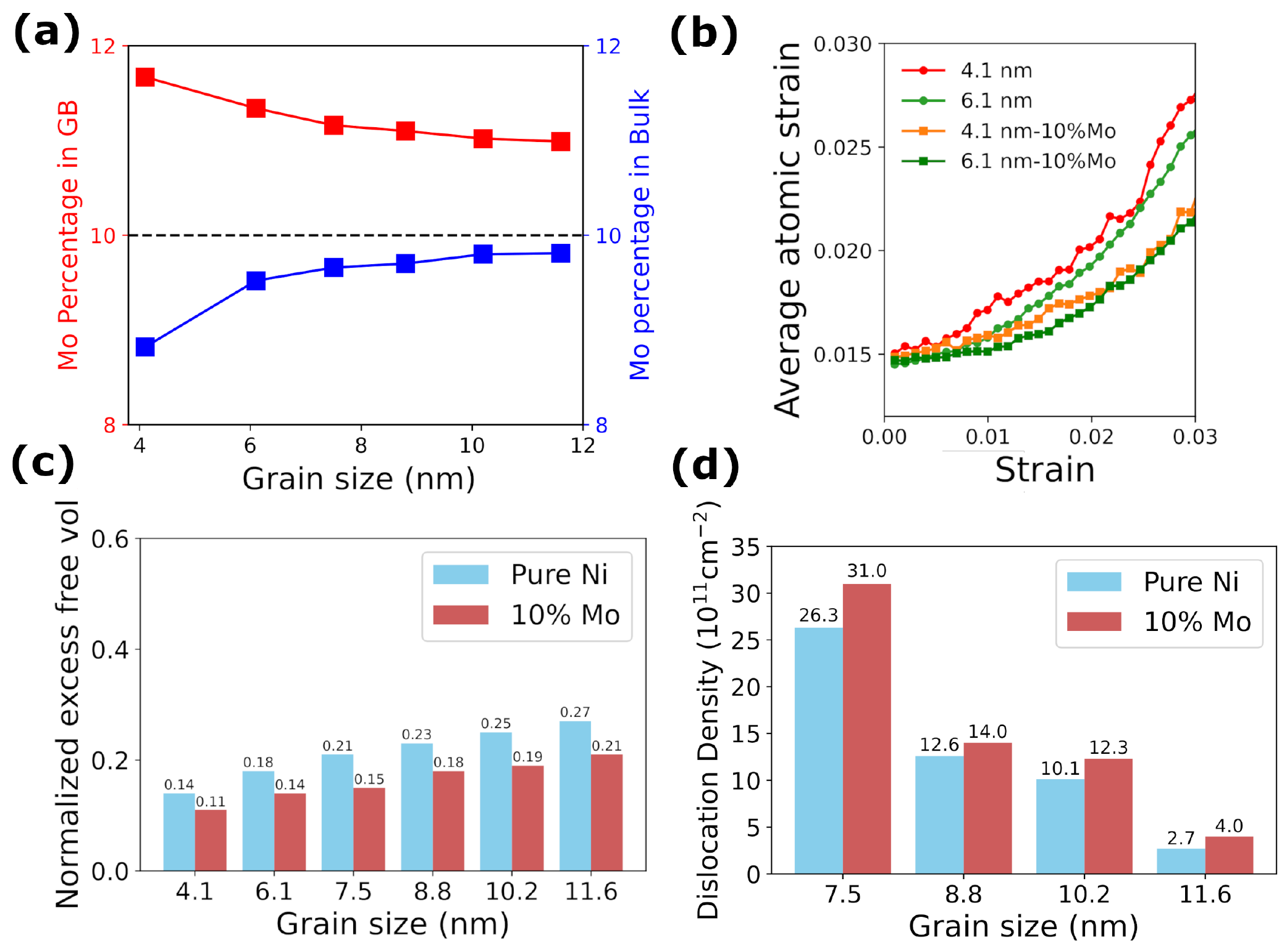}
\caption{\label{fig:doping}GB analysis of the polycrystal model. (a) Mo percentage in GB as a function of grain size. (b) The average atomic strain as a function of tensile strain for small grain sizes w/ and w/o doping of Mo. (c) Normalized excess free volume, and (d) dislocation density for different grain sizes w/ and w/o doping of Mo at $9\%$ applied strain.}
\end{figure}

We further investigated the solute effects on the mechanical properties of the polycrystals by replacing 10\% of all Ni atoms with Mo. As shown by the red points in Fig.~\ref{fig:hallpetch}b, the flow stress increases by a significant amount for all ranges of grain sizes, compared with pure Ni. To find out the underlying strengthening mechanism, we first analyze the segregation of Mo atoms. After MC/MD simulations, solute Mo atoms segregate at GBs, as shown in Fig.~\ref{fig:doping}a. This segregation effect becomes even stronger as the grain size decreases. In the meanwhile there still exists a significant amount of Mo atoms in the bulk region, which is consistent with the experimental results \cite{hu2017grain}. We further compare the average atomic strain of the doped systems with those without Mo doping in small grain size systems, as shown in Fig.~\ref{fig:doping}b. It is clear that after doping 10$\%$ Mo, the atomic strain at yielding is significantly reduced for grain sizes of 4.1 nm and 6.1 nm. This doping effect is weakened as the grain size increases (see Fig. S1 in the supplementary information). Therefore, for the small grain-size polycrystals when the GB mediated plasticity dominates, solute doping can greatly reduce the atomic strain leading to an increased flow stress. The calculations of normalized excess free volume in the GB also support this conclusion. We define the normalized excess free volume (denoted as $k$) as below,
\begin{equation}
\begin{aligned}
k & = (V_\mathrm{GB}-N_\mathrm{Ni,GB}\times v_\mathrm{Ni}-N_\mathrm{Mo,GB}\times v_\mathrm{Mo})/(N_\mathrm{Ni,GB}\times v_\mathrm{Ni}+N_\mathrm{Mo,GB}\times v_\mathrm{Mo}) \label{eq1}
\end{aligned}
\end{equation}
where $V_\mathrm{GB}$ is the total volume of GB atoms; $N_\mathrm{Ni,GB}$ and $N_\mathrm{Mo,GB}$ are numbers of Ni and Mo atoms in GBs, respectively; $v_\mathrm{Ni}$ and $v_\mathrm{Mo}$are the volumes of Ni and Mo atoms in the respective bulk. We use the common neighbor analysis algorithm in OVITO \cite{stukowskiVisualizationAnalysisAtomistic2009} to identify the GB atoms. In principle, GBs are less dense than a perfect crystal leading to a positive value of $k$. As shown in Fig.~\ref{fig:doping}c, the excess volume is reduced after Mo solute doping for different grain sizes. This means that the GB is denser after 10$\%$ Mo doping, making it more difficult for GB atoms jumping and free volume migrations. As a result, the resistance to GB-mediated plasticity increases, which will benefit for the flow stress increasing.

For large grain size polycrystals, the dislocation-based deformation becomes much more important. We thus plot the intragranular dislocation density for large grain sizes, as shown in Fig.~\ref{fig:doping}d. We can observe that dislocation density increases after Mo doping for large grain sizes. The increase of the dislocation density will in general result in the increase of the strength due to dislocation interactions and entanglements. This explains that the flow stress increases after Mo doping for polycrystals with large grain sizes.

\subsection{Annealing effect}

\begin{figure}[t]
\includegraphics[width=1.0 \textwidth]{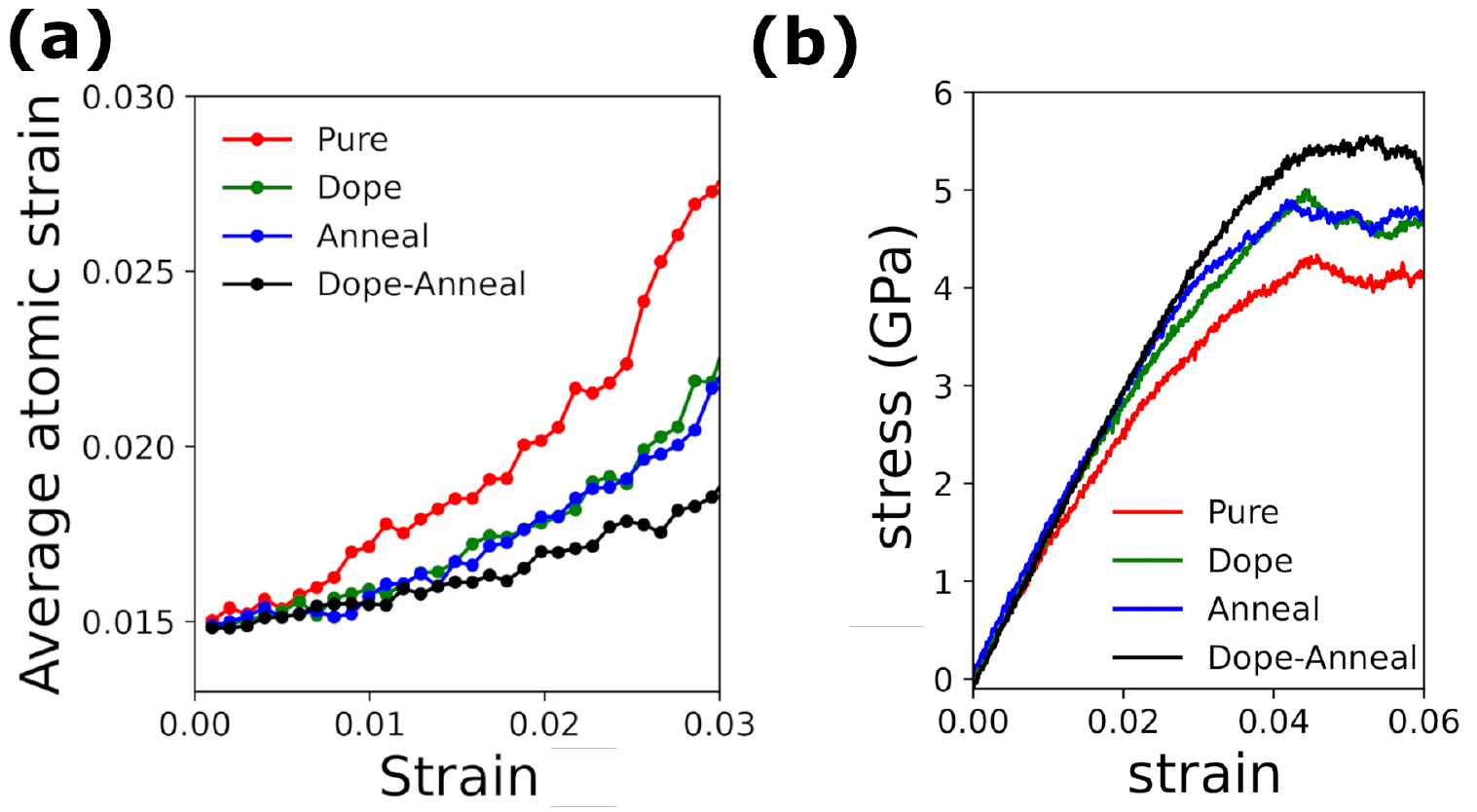}
\caption{\label{fig:anneal}Annealing effects on small-size polycrystal model. (a) The average atomic strain at yielding as a function of tensile strain under different conditions for polycrystal with grain diameter 4.1 nm. (b) The stress-strain curves under different conditions for polycrystal with grain diameter 4.1 nm. Pure: as-prepared Ni polycrystal after MC/MD calculations at room temperature; Doping: Doping the Ni polycrystal with 10$\%$ Mo; Annealing: anneal the pure Ni polycrystal at 600 K; Doping-Annealing: anneal the 10$\%$ Mo doped polycrystal at 600 K.}
\end{figure}

To study the annealing effect on the mechanical properties of the polycrystal, we annealed both the pure Ni and the doped with 10$\%$ Mo polycrystals with small grain size (e.g. grain diameter equals 4.1 nm) at 600 K. We calculated the average atomic strain for all four polycrystals, pure Ni (Pure), Ni with 10$\%$ Mo doping (Doping), pure Ni after annealing at 600 K (Annealing), and Ni with 10$\%$ Mo doping after annealing at 600 K (Doping-Annealing). As shown in Fig.~\ref{fig:anneal}a, doping or annealing alone will reduce the atomic strain at yielding with a considerable amount. With both doping and annealing, the atomic strain will reduce further compared with doping and annealing alone (also see Fig. S2 in the supplementary information for the strain distributions). The strain energy (defined as (potential energy-energy in the corresponding bulk)/number of atoms) of the polycrystals is 0.154 eV/atom for pure Ni, reduced to 0.145 eV/atom after aneealing and 0.122 eV/atom after Mo doping, and further reduced to 0.112 eV/atom with both doping and annealing. In other word, both doping and annealing can stabilize the GB, making the GB mediated plasticity more difficult. This can also be observed from the stress-strain curves in Fig.~\ref{fig:anneal}b. The plasticity occurs in pure Ni polycrystal much earlier (red curve), followed by the doping or annealing alone polycrystals, (green and blue curves); while the polycrystal with both doping and annealing has the stress increasing linearly for a much wider range of strain, resulting in the largest strength compared to other three polycrystals. Therefore, with both doping and annealing, the strength of the small-size polycrystals can be improved further, e.g., grain diameter small than 7 nm (see blue points in Fig.~\ref{fig:hallpetch}b).

\begin{figure}[h]
\includegraphics[width=1.0 \textwidth]{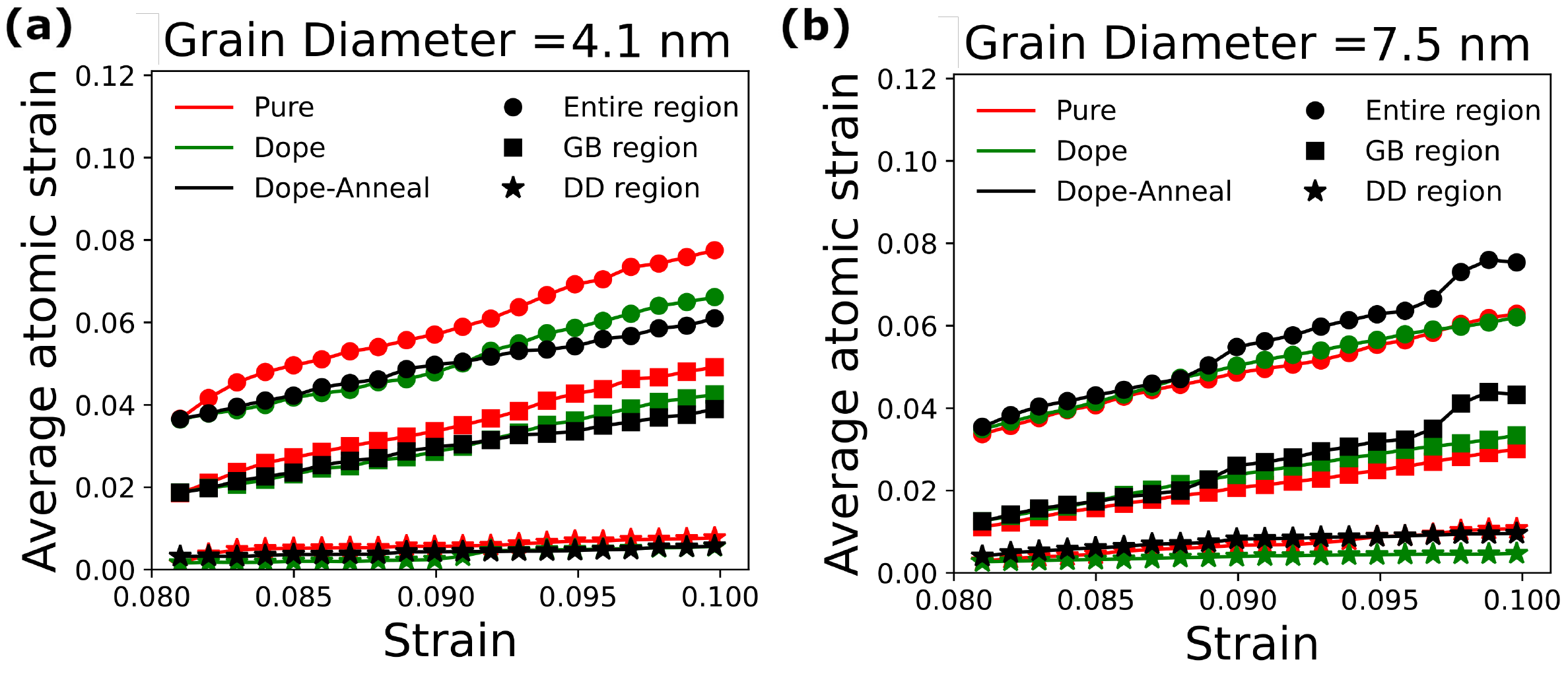}
\caption{\label{fig:anneal1}Annealing analysis of the polycrystal models with different grain sizes. The average atomic strain during plastic flow deformation as a function of tensile strain under different conditions for polycrystal with grain diameters (a) 4.1 nm, and (b) 7.5 nm, at entire, GB, and dissociated dislocation (DD) regions, respectively.}
\end{figure}

On the other hand, annealing has a opposite effect on large-size polycrystals, e.g., grain diameter larger than 7 nm, as shown by blue points in Fig.~\ref{fig:hallpetch}b). In other words, further annealing after doping will reduce the strength of the large-size polycrystals. To reveal the underlying atomistic behavior, we plotted the average atomic strain during plastic flow deformation for both small-size (grain diameter equals 4.1 nm) and large-size (grain diameter equals 7.5 nm) polycrystals, as shown in Fig.~\ref{fig:anneal1}. It is clearly observed that the atomic strain during plastic deformation can be further reduced after annealing for small-size polycrystal, while for large-size polycrystal annealing will increase the atomic strain significantly, mainly at GB region (see Fig.~S3 in the supplymentary information for the atomic strain distribution in the model structure).

\section{Discussion}

ML-IAP has received plenty of interest in the field of materials science due to its high accuracy and good scalability \cite{mishin2021machine}. It has been used to investigate a variety of mechanical properties in metal and alloys, including the dislocation properties \cite{yin2021atomistic,zhao2021anomalous,byggmastar2021modeling}, stacking faults \cite{bartok2018machine,wang2021generalized,li2020complex}, phase transition \cite{zong2018developing,verdi2021thermal}, etc. However, few studies were dedicated to the GB strengthening problem, e.g., different grain sizes in polycrystals. This is mainly due to the high computational cost of ML-IAPs compared to the classical force fields, e.g., EAM potential \cite{li2018quantum}. Here, with the help of high performance computing clusters, We successfully study polycrystals using ML-IAP because this type of IAP, like the classical ones, still scales linearly with the number of atoms and is orders of magnitude computationally cheaper than DFT calculations. We demonstrated that the ML-IAP can successfully reproduce the Hall-Petch and inverse Hall-Petch relations in polycrytalline metals and reveal their underlying mechanisms. 

Complex alloy strengthening mechanisms have been extensively explored, including the GB strengthening, solute strengthening, annealing strengthening, etc. GB strengthening, also known as Hall-Petch strengthening, will fail at extremely fine grain sizes, e.g., nanometer sizes. While after coupling with solute effects, the Hall-Petch maximum strengthening limit can be modified to lower grain size \cite{sansoz2022hall}. Annealing adds extra complications to the strengthening mechanisms, which will further strengthen the alloys for small grain size polycrystals while weakening the large-size ones by triggering large atomic movements in GB region during plastic deformation (see Fig.~S3 in the supplymentary information). The enhancement of the strength in small grain size polycrystals with solute doping and annealing can lead to the resurgence of Hall-Petch strengthening down to even smaller grain sizes, as shown in Fig.~\ref{fig:hallpetch}b (dashed black line).  

\section{Conclusion}
The present computational and theoretical study has shown that the Hall-Petch and inverse Hall-Petch relations can be well predicted by the ML-IAP. A switch of the dominant mechanism from the dislocation based to GB mediated plasticity is observed as the grain size decreases, which is responsible for the Hall-Petch to inverse Hall-Petch transition. Atomistic analysis indicates that uniaxial tensile loading induced atomic strain can be significantly reduced by both solute doping and annealing for small grain size polycrystals, resulting in the great enhancement in the strength of the polycrystals. However, for the doped large-size polycrystals, annealing would weaken the samples due to the increased atomic strain in GB region. The combination effects of solute doping and annealing on different grain size of polycrystals can postpone the inverse Hall-Petch relation and keep continuous strengthening down to even smaller grain sizes. Our atomistic simulations reveal the underlying complex strengthening mechanisms in nano-sized polycrystals, theoretically supporting the potential of achieving ultra-strong nanograined materials.

\section*{Acknowledgements}
XL, QZ, SL, and JS would like to acknowledge financial support from the Hundreds of Talents Program of Sun Yat-sen University.


\end{document}